# Negative thermal expansion of nanoporous anodic aluminum oxide membranes.

L. Forzani[1], C. A. Ramos[2], E. Vassallo Brigneti[3], A. M. Gennaro[1,4], and R.R. Koropecki[1]

[1]*Instituto de Física del Litoral, UNL-CONICET. Güemes 3450, 3000 Santa Fe, Argentina.*

[2]*Centro Atómico Bariloche, Comisión Nacional de Energía Atómica, Av. E. Bustillo 9500, R8402AGP, S. C. de Bariloche, Río Negro, Argentina.*

[3]*Departamento de Matemáticas y Física, Instituto Tecnológico y de Estudios Superiores de Occidente, ITESO, Av. Periférico Sur Manuel Gómez Morín 8585, C. P. 45604, Tlaquepaque, Jalisco, México.*

[4]*Departamento de Física, Facultad de Bioquímica y Ciencias Biológicas, UNL. Ciudad Universitaria, 3000 Santa Fe, Argentina*

**Abstract.**

We have measured the thermal expansion of Ni nanowires electrodeposited into self-organized nanoporous amorphous aluminum oxide (AAO) membranes without Al substrate using X-ray diffraction between 110K and 350K. The results indicate an average thermal expansion of the Ni nanowires -along the wire axis- of $\overline{\alpha}_{NiNW} = -(1.6 \pm 1.5) \times 10^{-6} \text{ K}^{-1}$. Assuming a bulk-like thermal expansion of the isolated Ni nanowires, this result indicates that AAO has also a negative thermal expansion. We estimate the thermal expansion of nanoporous AAO to be $\alpha_{AAO} = -(5 \pm 1) \times 10^{-6} \text{K}^{-1}$. We show that data obtained previously on the thermal expansion of metallic nanowires grown in the nanoporous AAO may be interpreted as originated in a negative thermal expansion of the matrix.

# Introduction

It is well known that nanostructured materials are characterized by different properties as compared with their bulk counterparts. Characterizing these properties is basic in research and applications [1]. One of the self-organized nanostructured materials most used is the nanoporous anodized aluminum oxide (AAO) [2, 3]. To characterize this particular nanomaterial it is necessary to determine its properties, such as elastic constants [4,5], annealing effects [6], thermal conductivity [7], thermal expansion [8], and Poisson ratio [9] among others. In this work we focus on an experimental estimate of the thermal expansion of nanoporous AAO using electrodeposited Ni nanowires as strain sensors.

Zhang *et al*. [8] made a significant effort to measure the thermal expansion of AAO using a modified AFM microscope to determine the thickness change of a 3 μm thick membrane of porous AAO grown on a glass substrate. The average $\alpha(T)$, between 300 K and 400 K, turns out to be about three times larger than the values obtained for bulk alumina [3]. Also, these values show an increase of ≈70% on going from 300 to 400 K, which may be compared with the bulk alumina change of only 3% increase in this temperature range [10].

Other authors have used AAO as templates to electrodeposite nanowires (NWs) of different metals and *in-situ* XRD to measure the thermal expansion of these embedded NWs. Among these, Xu *et al*. [11] measured a near-zero thermal expansion of Ag NWs (considering it from 0 to 650°C). In their analysis Xu *et al*. did not take into consideration the AAO matrix in which the Ag NWs were electrodeposited. Instead, they suggest vacancies in the Ag NWs were responsible for the observed $\alpha \cong 0$ for the NWs, in spite of not reaching the expected bulk value of $\alpha(T)$ of silver even after the samples were annealed up to 800°C. A similar result was reported when studying Cu NWs electrodeposited into the AAO templates by Zhou *et al.* [12]. Cai *et al*. [13] studied Ni NWs by in-situ XRD and EXAFS. By XRD they measure a thermal expansion coefficient similar to that of bulk Ni, but EXAFS gave a larger value. They explained the results proposing the presence of a 50% of amorphous Ni in their NW. It should be remarked that in [13] the Ni NW were electrodeposited by DC.

The large values for the thermal expansion coefficient of AAO reported by Zhang *et al*. [8] are not consistent with previously reported magnetic anisotropy change observed in Ni NWs grown into the nanopores of AAO, which points out to the fact that AAO has a thermal expansion coefficient *smaller* than Ni below room temperature [14-17].

In this work we report the thermal expansion coefficient of Ni NW embedded in the AAO matrix between 110K and 350 K. We argue that the measured value is mainly determined by the AAO matrix and gives an estimate of the as-prepared nanoporous AAO thermal expansion in this temperature range.

## 2. Experimental

Nanoporous AAO was produced by a two-step anodization process [2] using oxalic acid 0.3 M at 7°C and 40 V. The voltage after the second anodization was decreased exponentially from 40 V to 8.3 V in order to decrease the thickness of the insulating barrier layer between the nanoporous AAO and the Al substrate to the optimum value of 10 nm [18].

The Ni filling of the porous structure was made by pulsed electrodeposition using a repeated application of a negative pulse followed by a positive pulse to discharge the capacitor formed by the barrier layer [19].The cycle ends with a "dead time" at 0 V for rearrangement of the electrolyte into the pores. Using SEM we obtained an estimate of the Ni NW diameter, $d \cong 40$ nm and an average separation between the wire centers, $D \cong 120$ nm. Thus, the area-filling-factor, $f = \frac{\pi}{2\sqrt{3}}\left(\frac{d}{D}\right)^2$, of the Ni NW with respect to the total area is approximately 10% [20].

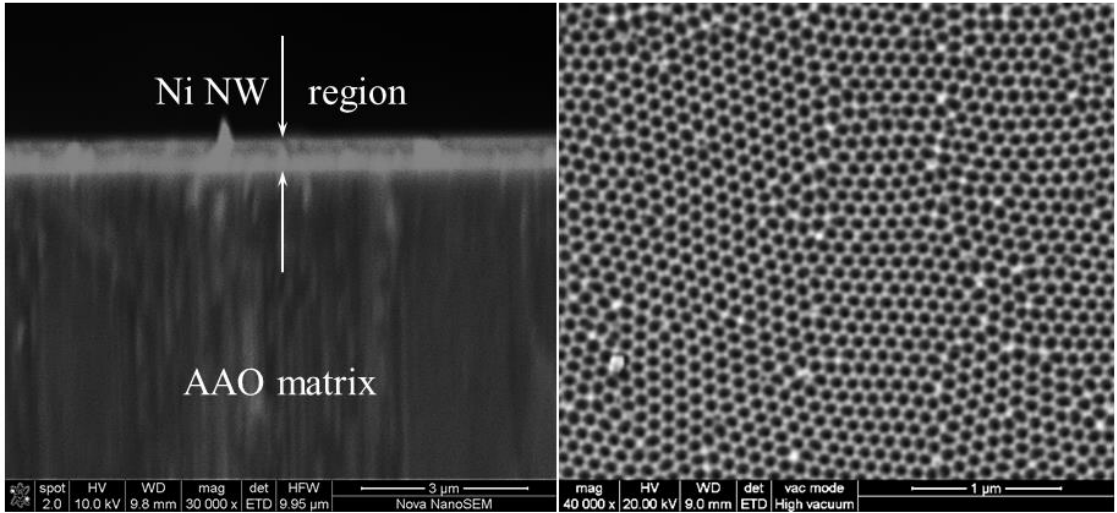

Fig. 1. Left: Lateral view of one of the samples used in this study showing the length of the nanowires. Right: Top view of the same sample. The pores are ≈ 40 nm in diameter and their center-to-center spacing is ≈ 120 nm, indicating a filling factor, $f \cong 0.1$.

The resulting NW length depends on the electrodeposition conditions and time. In the samples considered here, the Ni NW are much shorter than the AAO thickness. Figure 1 shows a lateral and a top view of one of the samples under study.

The aluminum substrate was removed by chemical etching with a HCl-CuCl$_2$ solution. Thus the samples under study, typically ∼ 20μm thick are not subject to the large thermal contraction of the Al substrate.

X-Ray diffraction was performed on a θ-2θ configuration in a PW1710 Philips with a Cu K$_\alpha$ line. A commercial low temperature controller, PAAR TTK-450, was used to fix the temperature, $T$, in the range 110 K ≤ $T$ ≤ 350 K. The samples were set in thermal contact with a Si (001) wafer by a small amount of vacuum grease covering less than ≈ 5% of the sample area to decrease possible strain effects on the membrane. After the measurement was performed the sample was removed unbroken. The Si wafer was used with a double purpose: to eliminate background from the sample holder and to provide with an in-situ thermal expansion reference when the Si(004) reflection condition was met. The sample height was adjusted to be at the required X-ray beam position by properly adjusting a Cu supplement under the Si wafer. The Si (004) reflection could be eliminated by tilting slightly the sample angular position, which could be adjusted independently of the angular position of the detector (2θ). We checked that the Ni NW $\theta - 2\theta$ scans were not affected this small tilt ($< 1°$). In figure 2 we show a $\theta - 2\theta$ XRD scan from which we deduced that the AAO membrane gives broad peaks associated to its amorphous structure while the Ni NW show the expected FCC peaks with a lattice parameter (352.5 ± 0.4) pm, which matches the reported bulk value. Applying Scherrer´s formula to the (111) and (220) peak profiles (figures 3 and Supplementary Material) we estimate a crystallite size of 25 and 75 nm respectively. These magnitudes are similar to what other authors have reported [13].

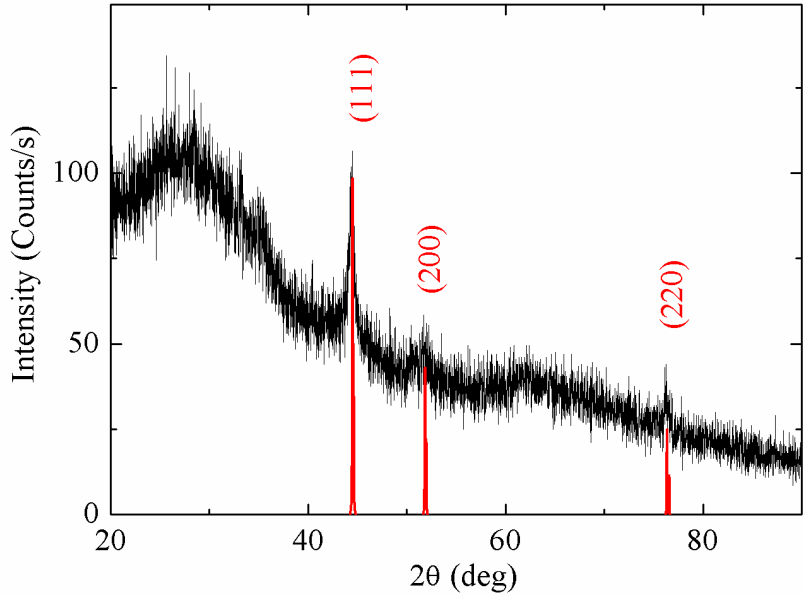

Fig. 2: Room temperature XRD of the Ni nanowires embedded in the porous anodic aluminum oxide template. The Ni peak positions coincide with the expected bulk values. The anodic aluminum oxide matrix shows broad peaks consistent with its amorphous character.

## 3. Thermal Expansion results

In figure 3 we present the $\theta - 2\theta$ scans of a sample around the (111) reflection at two different temperatures, $T = 125K$ and $T = 310K$. The peak line shape was adjusted using non-linear fitting routine with two of Lorentzian lineshapes of intensity 2:1 associated to the CuK$_{\alpha 1}$ and Cu K$_{\alpha 2}$ lines plus a linear background. In Figure 3 we indicate the peak position. The (111) reflection at $T = 310K$ is shifted for clarity. The line position was determined within an uncertainty of 0.04°. A similar behavior was obtained from the (220) reflection (Supplementary Material). The temperature was cycled from room temperature to $T \cong 120K$ and back each time we measured around the Ni (220), Si (004), and finally Ni(111) peaks. Thus we claim the observed behavior is reproducible.

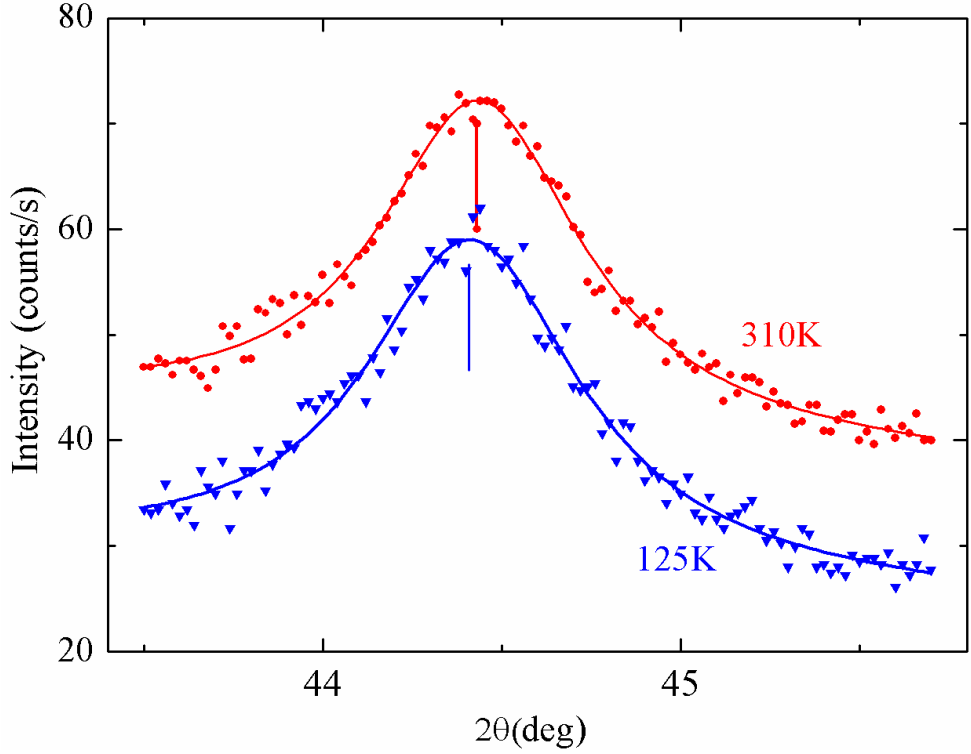

Figure 3: XRD scan of the (111) Ni NW reflection. The solid line corresponds to the best fit of the data to two Lorentzian lineshapes in a 2:1 intensity ratio as expected from the $K_{\alpha 1 : }K_{\alpha 2}$ and a linear background. The 310K line was shifted upward for clarity.

In figure 4 we plot the (111) peak positions as a function of temperature. Performing a linear regression of the data between 125K and 350K we obtain $2\theta_{111} = (44.405 \pm 0.006)\text{deg} + (1.04 \pm 0.26) \times 10^{-4}(T(K) - 300K)$ deg K$^{-1}$. The positive slope in $2\theta_{111}(d\theta/dT = (0.91 \pm 0.23) \times 10^{-6}$rad/K) indicates a small negative thermal expansion coefficient given by:

$$\alpha(T) = -\frac{1}{tan\theta}\frac{d\theta}{dT} \qquad (1)$$

from where we deduce an average thermal expansion coefficient between 125K and 350K of $\bar{\alpha}_{NiNW} = -(2.2 \pm 0.6) \times 10^{-6} K^{-1}$ for the Ni NW embedded in the AAO matrix.

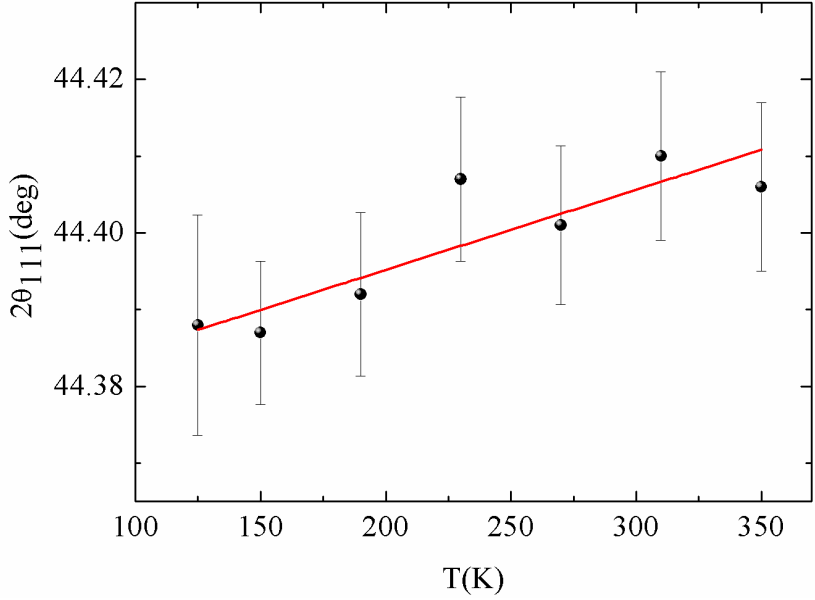

Figure 4: linear fit of the (111) peak position as a function of *T*. Note that this positive slope indicates a negative thermal expansion of the Ni NW embedded in the AAO matrix.

The Ni (220) Bragg reflection suggests a similar result: $\bar{\alpha}_{NiNW} = -(2 \pm 3) \times 10^{-6} K^{-1}$ (Supplementary Material), albeit with a larger error. We have considered the effect of systematic errors derived from: a) sample height changes, and b) differences between the sample temperature and the programed temperature (Supplementary Material). From our experiments we conclude that $\bar{\alpha}_{NiNW} = -(1.6 \pm 1.5) \times 10^{-6} K^{-1}$ in the temperature range of this study.

## 4. Discussion

The average thermal expansion coefficient of *bulk* Ni between 125K and 300 K is $\bar{\alpha}_{Ni} = 11.4 \times 10^{-6} K^{-1}$ [21]. For bulk alumina, using the results of Hayashi *et al.* [10], $\bar{\alpha}_{Al2O3} = 4.0 \times 10^{-6} K^{-1}$. It is obvious that the Ni NW thermal expansion does not behave either as bulk Ni or as bulk alumina.

There is an additional clue as to what is happening to the Ni NW provided by the magnetic anisotropy associated to the large magnetostriction of this material [14] [16] [22]. When studying the magnetic behavior of Ni NW embedded in a nanoporous AAO matrix these authors observed a reduced magnetic anisotropy upon cooling, which is consistent with the anomalous thermal expansion of the Ni NW in the AAO matrix associated with an *elongation* of the NW upon cooling. However, in these works the Al substrate was not removed, which has been claimed to be responsible for significant magnetoelastic effects observed. We note that, if the strain parallel and perpendicular to the NW axis were equal, then the magnetostrictive effects on the Ni NW would compensate. A more detailed study of the magnetostriction effect on the magnetic properties of this system is under way. Magnetoelastic effects are a consequence of the anomalous strains of the Ni NW in the matrix, the latter being caused by the mismatch between the NW and the matrix thermal expansion and elastic properties as we suggest below.

If a perfect bond is assumed between the Ni NW and the AAO matrix, the measured thermal expansion coefficient would correspond to that of the nanocomposite. When considering the mechanical response of aligned fiber composites, Mallick [23] shows that the composite average thermal expansion can be written as:

$$\bar{\alpha}_c = \frac{fE_{Ni}\bar{\alpha}_{Ni} + (1-f)E_{AAO}\bar{\alpha}_{AAO}}{fE_{Ni} + (1-f)E_{AAO}} \qquad (2)$$

where $f$ is the filling factor of the Ni nanowires, $E_{Ni}$ and $\bar{\alpha}_{Ni}$ are the Young modulus and average thermal expansion coefficient of Ni and similarly for the nanoporous AAO. The thermal expansion measured corresponds to the composite under the assumption of elastic deformation and a perfect bond between the Ni nanowires and the AAO matrix ($\bar{\alpha}_c = \bar{\alpha}_{NiNW}$). Similar result as Eq (2) was used more recently by Piraux *et al.* when considering polycarbonate embedded magnetic nanowires and magnetoelastic effects [24]. Clearly, if $f \to 0$ then the measured thermal expansion of the composite would correspond to that of the AAO matrix. To account for the Ni modification of the observed thermal expansion of the composite as compared to the AAO thermal expansion we simply convert equation (2) into:

$$\bar{\alpha}_{AAO} = \bar{\alpha}_{NiNW}(1+C) - C\bar{\alpha}_{Ni} \qquad (3)$$

where $C = [E_{Ni}f/E_{AAO}(1-f)]$. The Young moduli of Ni and as-prepared nanoporous AAO are $E_{Ni}$= 200GPa [25], and $E_{AAO} = 147$GPa [4] or $E_{AAO} = 114$GPa [5], from which we obtain an average of $C = (0.17 \pm 0.04)$ where the error considered amounts to a rough 20% in $C$. From Eq. (3) we obtain $\bar{\alpha}_{AAO} = -(3.8 \pm 1.9) \times 10^{-6} K^{-1}$ for the average thermal expansion of the unperturbed nanoporous AAO matrix along the pore direction. This negative value for the thermal expansion of AAO contrasts with the bulk alumina value which is positive, yet of similar magnitude.

Searching for previously reported thermal expansion of other NW we found the results of Xu *et* al. on Ag NWs grown on AAO [11]. Xu *et al.* report an average thermal expansion of as-prepared Ag NW of $\bar{\alpha}_{AgNW} = 6.35 \times 10^{-9} K^{-1}$ between room temperature and 650°C. Note that this value is more than three orders of magnitude smaller than bulk Ag: $\bar{\alpha}_{Ag} = 20.8 \times 10^{-6} K^{-1}$ for the average value between room temperature and 571°C [26]. From the reported preparation conditions, which yield a definite pore distance *D*, and the reported pore diameter *d*, we calculate their filling factor $f = 0.27 \pm 0.05$. (see Supplementary Material). Considering a reported Young modulus for Ag of $E_{Ag} = 85$ GPa [27] we derive $C = 0.24 \pm 0.06$. From these results we obtain a corrected thermal expansion for the unperturbed AAO matrix of $\bar{\alpha}_{AAO} = -(5.0 \pm 1.3) \times 10^{-6} K^{-1}$.

When measuring by X-ray diffraction above room temperature in Cu NW electrodeposited in to AAO, Zhou *et al*[12] report a nearly zero thermal expansion between room temperature and 300°C: for the (220) Bragg diffraction peak using their Figure 4.c we obtain $\bar{\alpha}_{CuNW} = (0.0 \pm 0.5) \times 10^{-6} K^{-1}$ for the as-prepared samples. In their text Zhou *et al* show that the Cu NW diameter is 30nm. From their AAO preparation conditions, we calculate $f = 0.23 \pm 0.05$. Considering $E_{Cu} = 130$ GPa [27] we obtain: $C = 0.30 \pm 0.06$. The reported bulk Cu length change between 577 and 293K is $\Delta l/l = 0.005$ which would yield an average $\bar{\alpha}_{Cu} = 17.6 \times 10^{-6} K^{-1}$ [26]. From this we derive $\bar{\alpha}_{AAO} = -(5.3 \pm 1.1) \times 10^{-6} K^{-1}$.

In the same way, for Fe NWs in AAO matrix Xu *et al*[28] show a slightly negative thermal expansion $\bar{\alpha}_{FeNW} \approx -0.2 \times 10^{-6} K^{-1}$ for the data between room temperature and 250°C. From their work we estimate $f = 0.19 \pm 0.04$, $C = 0.34 \pm 0.07$, which yield $\bar{\alpha}_{AAO} = -(5.1 \pm 1.0) \times 10^{-6} K^{-1}$.

The above arguments assume that NWs behave as bulk material regarding its thermal expansion and elastic properties. When going to sizes comparable to the lattice parameter in one [29], two [22] or three dimensions [30] we may expect these properties to change following finite-size-scaling [31, 32]. We noted that our samples crystallize in the FCC structure with a lattice parameter consistent with bulk values. We also noted that the NW crystallite size is ~ 100 lattice parameters. The departure from bulk values for this NW diameter is expected to be very small. Indeed we were not able to differentiate the lattice parameter from its bulk value.

Recently Ho *et al*. [33] calculated the thermal expansion of isolated ultrathin NW of several FCC metals using molecular dynamics. Although they found negative thermal expansion for some of the metals studied, Ni NW of diameter $d \cong 5$ nm show a positive thermal expansion coefficient slightly less than the bulk value, and it is expected to approach the bulk value following the finite-size scaling hypothesis.

The magnitude of the finite-size-effect on the magnetic transition temperature of Ni NWs as a function of their diameter was reported to be ~ 1% (decrease) for $d = 40$nm [22]. We expect a similar order-of-magnitude effect on the thermal expansion coefficient of isolated Ni NW of similar diameter.

Nonetheless, an independent measurement of the thermal expansion of isolated Ni NW and of the nanoporous AAO membranes are necessary to corroborate the results deduced in this work using Ni NW as deformation sensors embedded in AAO.

In conclusion, our results on the thermal expansion of the Ni NW embedded in a nanoporous AAO matrix from 110K up to 350K, as well as previous results on Ag-, Cu- and Fe-NWs above room temperature, can be interpreted as associated to a *negative* thermal expansion coefficient of AAO along the pore direction with an average value of $\bar{\alpha}_{AAO} = -(5 \pm 1) \times 10^{-6}$ K$^{-1}$. The origin of the negative thermal expansion pointed by this study is certainly intriguing, and it may be linked to perpendicular motions of atoms or groups of atoms arranged differently from the crystalline state [34].

**5. Supplementary Material:** Results of thermal expansion measured on the Ni (220) Bragg reflection, filling factor determination from anodizing conditions, and systematic error considerations due to sample height and temperature difference between the programed temperature and sample temperature.

**Acknowledgements:** L.F. has a scholarship from CONICET. A.M.G. and R.R.K. are members of the CIC-CONICET. R.R.K. acknowledges support from PICT grant 2014-1683.C.A.R. thanks ITESO invitation as visiting Professor (June 2018), PICT 2015#883 and SePCyT UNCuyo 06/C263.

# Supplementary Material: "Negative thermal expansion of nanoporous anodic aluminum oxide membranes"

*L. Forzani, C. A. Ramos, E. C. VassalloBrigneti, A. M. Gennaro, and R. R. Koropecki.*

### S1) Thermal expansion measured on the Ni (220) peak

The peak (220) of Ni also shows a negative thermal expansion. This can be observed in the scan at two different temperatures, T = 300K and T = 110K in the figure below.

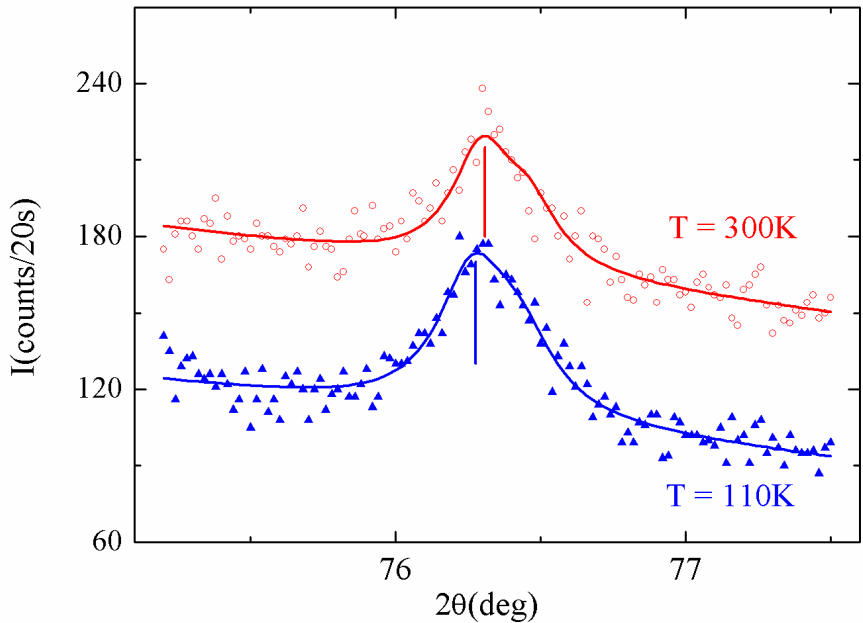

Figure S1: $\theta - 2\theta$ scan of the same sample near the (220) Bragg diffraction of Ni nanowires embedded in AAO at T = 300K and 110K. Note a slight shift towards lower angles at low temperature, indicatingalarger lattice spacing along the nanowire axis.

The peak profile was adjusted following a two Lorentzian peaks centered at $K_{\alpha 1}$ and $K_{\alpha 2}$ with 2:1 intensity ratio and the same width. The observed shift (of only $\Delta 2\theta = 0.032°$)traduces into a negative average thermal expansion $<\alpha_{(220)}> = -(2 \pm 3) \times 10^{-6} K^{-1}$, where the uncertainty derives from in the peak position error determination. The vertical lines indicate the best-fit curve maxima.

### S2 Filling factor and anodizing conditions

Nanoporous AAO is prepared in a two-step anodizing process from a high purity (99.999 %) aluminum foil under specific conditions that produce a self-organized structure of compact hexagonal cells with centered cylindrical pores perpendicular to the substrate. The filling factor $f$ or porosity parameter, representing the fraction of transversal area covered by the nanowires, is calculated as $f = \frac{\pi}{2\sqrt{3}}\left(\frac{d}{D}\right)^2$, where $d$ is the pore diameter and $D$ the inter-pore distance [1].

The relevant preparation conditions are the kind of electrolyte, the anodization voltage, and to a lesser extent the temperature of the process. Each electrolyte, together with the voltage, determine the parameters $d$ and $D$. As an example, pore diameters $d \sim 40$ nm and inter-pore distances $D \sim 105\text{-}120$ nm are obtained using oxalic acid 0.3 M at 40V, while $d \sim 20$ nm, and $D \sim 60$ nm are produced using sulfuric acid 0.3 M at 25 V [2].

After the anodization process it is possible to widen the pore diameter using a chemical etching with phosphoric acid 5% w/w at different times. This process does not change $D$, and makes it easier the deposition of metals inside the pores.

Besides our experimental data on Ni nanowires, we have analyzed the experimental results of other authors working with Cu-, Ag- and Fe-nanowires electrodeposited in AAO. Here follows the calculation of the filling factor, $f$, for each case:

- Ni NW (our data): AAO was prepared with oxalic acid 0.3 M, at 40V and 7°C. The samples were characterized by SEM and AFM, having $d = 40$ nm, and $D = 120$ nm, giving $f = 0.10$.
- Ag NW [3]: AAO was prepared using the process by Masuda and Satoh [4], using oxalic acid 0.3 M, 40 V at 17°C, which yields $D = 100$ nm. Xu *et al* [3] measured $d = 55$ nm, thus $f = 0.27$.
- Cu NW [5]: AAO was prepared with sulfuric acid 0.3 M, at 25V, which yields $D = 60$ nm. The authors report $d = 30$ nm, from where we estimate $f = 0.23$.
- Fe NW [6]: AAO was prepared with oxalic acid 0.3 M, at 40V, producing $D = 120$ nm. The reported $d = 55$ nm, from this we estimate $f = 0.19$.

A 10% variation in either $d$ or $D$ would lead to an estimated 20% uncertainty in $f$, which is what is considered in the main text.

**S3 Systematic errors considerations**

The effect of sample height modifications, as well as possible temperature ($T$) differences between the programed $T$ and the actual $T$ of the sample volume tested in the experiment, may occur with the variable-$T$ setup used in this work. In order to test the effect of the sample height effect on the observed peak position we measured the Si(004) peak at three $T$ around room $T$, where Si has a very small but positive thermal expansion of $+2.5 \times 10^{-6} K^{-1}$. The diffractograms are shown in Figure S2 left, together with the best fit using symmetric peaks (mixture of Lorentzian and Gaussian) with adjustable widths. The peak positions corresponding to $\lambda_{K\alpha 1} = 0.154059$nm obtained from the best fits were: $2\theta_{Si(004)} = (69.1018 \pm 0.0014)°$, $(69.1038 \pm 0.0013)°$, and $(69.1046 \pm 0.0012)°$ for $T = 350K$, 300K and 250K respectively, as programed temperatures. The observed displacement would indicate a positive thermal expansion of Si, yet much smaller than expected. Indeed, the peak positions would have had a significantly larger angular displacement to meet the measured thermal expansion of Lyon *et al.*[7] using a capacitance dilatometer. The relative lattice parameter change measured by XRD and Ref [7] are plotted in Fig. S2 right.

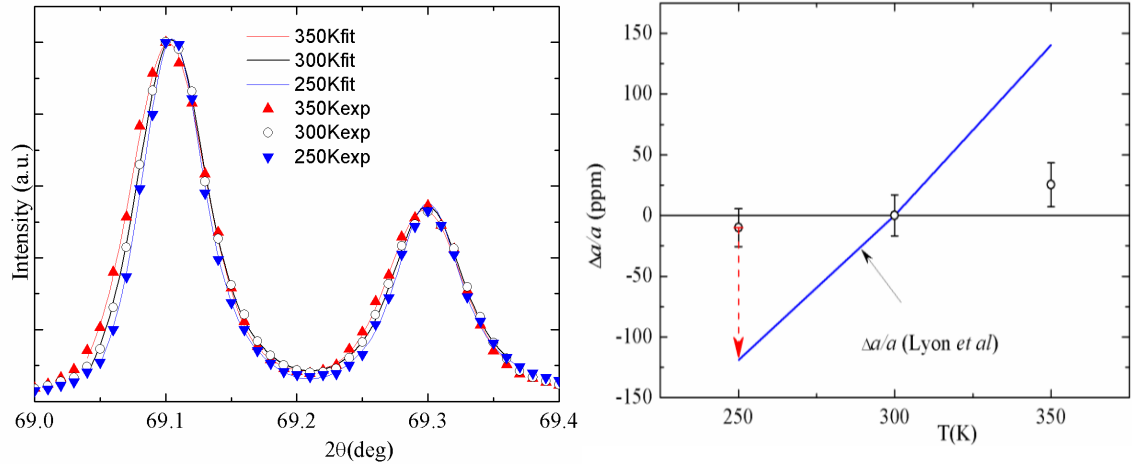

Fig S2: Left: XRD peak of the Si (001) single crystal used as sample holder immediately below the nanoporous AAO sample containing the Ni nanowires under study. The reduced scan shows the Si(004) peaks at three programed temperatures, 350K, 300K and 250K, and line profile fits, from where we extracted the best-fit position. Right: Relative lattice parameter change as a function of $T$, where 300 K was taken as reference. By XRD the deduced lattice parameter change of Si (hollow circles) is much smaller than that previously reported by Lyon *et al* [7] (solid line). The difference, indicated by a red dash-line may be attributed, mainly, to sample displacement effects.

The significant difference between the peak position and the reported thermal expansion results [7] are emphasized by a red arrow in Figure S2 right. This difference may be attributed, mainly, to sample height displacement with T. Indeed, Si, having such a low thermal expansion coefficient, is ideal to test sample height displacement. The predicted apparent lattice parameter change due to sample displacement is known as the Nelson-Riley function, given by [8]:

$$\frac{\Delta a}{a} = \frac{\delta \cos^2\theta}{R \sin\theta} \quad (S1)$$

where $\delta$ is the sample displacement and R is the diffractometer radius. The red dash-linein figure S2 right corresponds to $\Delta a/a = (1.1 \pm 0.2) \times 10^{-4}$ and, considering R = 200 mm, this would indicate a sample displacement of about $\delta = 18\mu m$. This correction is larger when considering the Ni(111) Bragg as the factor $\left(\frac{\cos^2\theta}{\sin\theta}\right)$ increases by 1.9 on going from $\theta_{Si(004)} \approx 34.5°$ to $\theta_{Ni(111)} \approx 22°$. If $\delta$ is considered to be constant over the approximately 200K temperature span (from 125K to 350K) then the correction of the effective thermal expansion coefficient would then be an *increase* of the reported value by $+1.0 \times 10^{-6} K^{-1}$. This correction increases the average thermal expansion determined to yield $<\alpha_{(111)}> = (-1.2 \pm 1.2) \times 10^{-6} K^{-1}$ [9]. The sample displace-ment correction applied to the Ni(220) peak ($2\theta = 76.2°$) is a factor 0.83 of the Si(004) peak, leading to a correction of $+0.4 \times 10^{-6} K^{-1}$, thus the resulting average thermal expansion determined in association with the $<\alpha_{(220)}> = (-1.6 \pm 3.0) \times 10^{-6} K^{-1}$ [9]. The uncertainty associated with this systematic correction was estimated to be of the same magnitude as the correction ($\pm 1 \times 10^{-6} K^{-1}$ for the (111) peak, and of $\pm 0.4 \times 10^{-6} K^{-1}$ associated to the (220) peak). Thus we conclude that the average thermal expansion of the Ni NW is $<\alpha> = (-1.3 \pm 1.3) \times 10^{-6} K^{-1}$ [9]. The actual sample *T*, on the other hand, can be different from the programed *T*. This would be best appreciated in materials with a well characterized and relatively large thermal expansion, and preferentially good thermal conductors. All these requirements are met by Al. In Figure S3 we show the lattice parameter change observed by XRD using the same set up to measure an ultrapure Al substrate.

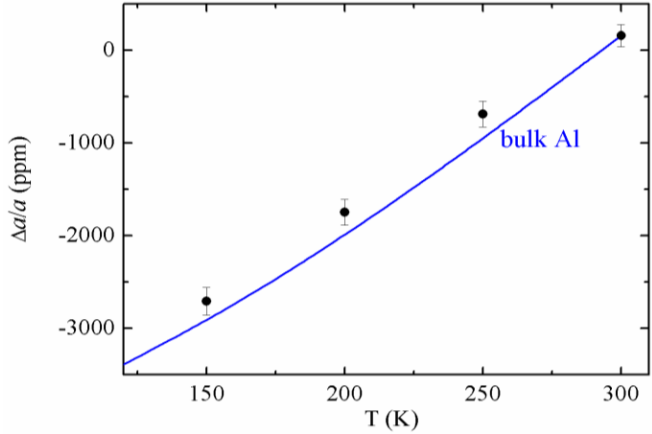

Fig. S3 Thermal expansion measured by XRD using the Al (311) peak and bulk value.

Here the sample positioning alone could explain the small departure from the bulk Al thermal expansion, but if we consider this departure to be due to the fact that the surface temperature of Al could be slightly *higher* than the programmed *T*, then the difference is $\Delta T = T_{surf} - T_{prog} = (12 \pm 6)K$. [E. Vassallo Brigneti *et al*, Ref. [9] of main text). In a 225K temperature span (from 125K up to 350K) this difference would make the temperature span 5% smaller and consequently a thermal expansion 5% larger. Silicon has a thermal conductivity of 150W/mK, slightly lower than the corresponding to Aluminum (235W/mK) [10], thus we would expect a slightly larger difference between the surface *T* of Si as compared with Al. Thermal conductivity of nanoporous AAO is a subject of current research [11]. Due to the fact that the resultant thermal expansion observed and corrected for sample shift is $<\alpha> = (-1.3 \pm 1.2) \times 10^{-6} K^{-1}$, a smaller temperature span (perhaps only 80%) of the programmed temperature span would only increase this result only by 25%. Considering *both* effects (sample height change and 80% smaller *T* span) we estimate the corrected average thermal expansion to be: $<\alpha> = (-1.6 \pm 1.5) \times 10^{-6} K^{-1}$.